\documentclass[12pt]{iopart}
\usepackage{graphicx}
\begin{document}

\title{Radial flow afterburner for event generators and the  baryon puzzle}

\author{E.  Cuautle  and  G. Paic}
\address{ Departamento de F\'{\i}sica de Altas Energ\'{\i}as\\
 Instituto de Ciencias Nucleares,\\
 Universidad Nacional Aut\'onoma de M\'exico, 
M\'exico D.F., C.P. 04510.}
\eads{\mailto{ecuautle@nucleares.unam.mx},\mailto{guypaic@nucleares.unam.mx}}

\begin{abstract}
A  simple   afterburner  to add  radial  flow   to  the  randomized
transverse  momentum  obtained from  event  generators, PYTHIA  and
HIJING,  has been  implemented  to calculate  the  $p/\pi$ ratios  and
compare  them with available  data.  A  coherent trend  of qualitative
agreement  has been obtained  in $pp$ collisions and in $Au+Au$
for various centralities. Those  results indicate that the radial flow
does play an important role in the so called baryon puzzle anomaly.

(Some figures in this article are in color only in the electronic version)

\end{abstract}

\submitto{J. Phys. G: Nucl. Part. Phys.}

\maketitle

\section{Introduction}

The study of  heavy ion  collisions at  the Relativistic  Heavy Ion
Collider (RHIC) has brought about very  strong evidence for the creation of a
very high energy density,  low baryon chemical potential, medium which
cannot simply be described in  terms of hadrons~\cite{exp}. Prominent
demonstrations  of this  new medium  created are:  parton  energy loss,
evidence for  very rapid thermalization  of the hot plasma  created in
collisions  from  the measurement  of  azimuthal  flow, and  abundant
production  of baryons  compared to  the  case of the some  yields observed  in
proton proton collisions.  The first effect we mention {\it i.e.} the parton
energy loss in the hot plasma results in a suppression of the yield of
high pt ($p_t > 4$ GeV/c) mesons by a factor of $\approx$ 5 compared to the
one measured in pp collisions.  This is quantified by the parameter $R_{AA}$
given by

\begin{equation}
R_{AA}= \frac{ (\frac{dN}{d^2p_Tdy})^{AA} }{T_{AA}(b) (\frac{dN}{d^2p_Tdy})^{pp} }
\end{equation}

\noindent
and/or the measurement of
the disappearance of ``away side'' jets. $T_AA(b)$ is the overlap
function between the two nucleons, as function of the impact
parameter, $b$. The measurement of the same ratio
for baryons ($p$, $\bar p$,$\Lambda$ and $\bar \Lambda$) have brought
a surprise:  the value  of $R_{AA}$ was  completely different from  the one
observed for  pions indicating a much lesser  apparent suppression or,
what  is more probable,  an increase  in the  baryon production  in the
range where the  excess is observed. This effect  was not predicted by
theory contrary to  the parton energy loss  and  the azimuthal flow, and
up to date  does not have a completely  satisfactory explanation. This
somewhat anomalous behavior of the baryon production is called the baryon
puzzle and refers to the  $p/\pi^+$ and $\bar p/\pi^-$ ratios measured
in the heavy ion collisions and even in pp collisions.
  
\noindent

The  studies of particle  production as  a function  of $p_t$,  in the
momentum regions where identification is possible at RHIC, exhibit the
following behavior:  the $p/\pi^+$ and $\bar  p/\pi^-$ ratios increase
with $p_t$  up to $\approx$ 2  GeV/$c$ and then start  to decrease for
higher       $p_t$      in      both       $pp$~\cite{star01}      and
$Au+Au$~\cite{star01,phenix,Star1} collisions, reaching a value
which  corresponds to  the  fragmentation value  observed in  $e^+e^-$
collisions for  quarks and  gluons~\cite{opal}.\\ The spectra  at $p_t
<2$  GeV/$c$ have  been observed  to follow  a  $m_t$~\cite{star2} and
$x_T$~\cite{star01} scaling, consistent with a transition between soft
and hard processes at around $p_t \approx 2$ GeV/$c$.

\noindent
The surprise  lies in the fact that  one would expect a  ratio that
does not exceed  the fragmentation value {\it i.e.} $\approx 0.2$ as observed in
$e^+ e^-$ collisions, while in the experiment the ratio rises up to more than one!\\
\noindent
In the  literature two  possible explanation are  prominently put
  forward:
\begin{itemize}
 \item the hydrodynamical approach~\cite{Hirano:2003pw,Peitzmann:2003ni,
Eskola05}   where  one   assumes  a   local  thermal   equilibrium  of
partonic/hadronic matter at an initial time, describing the space-time
evolution  of   thermalized  matter  by  solving   the  equations  for
energy-momentum  conservation  in the  hydro  picture. Another  model,
where the radial flow and the size of the system of emitting particles
are taking into account~\cite{ayala},  can describe the proton to pion
ratio for different centralities.\\
The hydrodynamical picture has been used to explain the broad features
of the increase in the baryon/meson ratio with limited success by Kolb
and Heinz~\cite{Kolb:2003dz}.

\item A large class of models called generically
"coalescence"  where  the  particle  species ratios  observed  in  the
intermediate  $p_T$ regime  (2-6 GeV/c)  of heavy  ion  collisions are
explained by  a collective production  mechanism, namely recombination
or  coalescence~\cite{Molnar:2005wf}.    In  most  coalescence  models
hadrons are  assumed to form  from essentially collinear  partons. The
parton  overlap function  is sometimes  simply assumed  to be  a delta
function, or at best in some cases small finite transverse widths have
been used, assuming  an $x_T$ distribution like one  expects to see in
the final state hadron, such that the partons do not have to undergo a
change in  momentum when forming  a hadron.  Although  the coalescence
models have been accepted, they do not provide a satisfactory response
to   many  questions~\cite{Peitzmann:2005ty}.
\end{itemize}
Furthermore in  our
opinion there also is  a fundamental contradiction between the fitting
of the spectra with the  coalescence approach which involves also some
flow contribution and the thermal analysis where, by a simultaneous fit
to the slopes of pions, kaons and protons one extracts the temperature
and the corresponding flow~\cite{NA44}. Since the coalescence approach
{\bf does  change} the  proton slope  with respect to  the one  of the
mesons even in absence  of flow, the temperature--flow analysis should
be reconsidered if the coalescence model is to be accepted.  Recently,
experimental  data measured  at two  different energies  also indicate
that  the  trend  of  the  coalescence  models  do  no  fit  the  data
\cite{Abelev:2007ra}\\ 
It is a fact that none of the models have been tested in a wide
centrality and energy range or with different projectiles, so  that
the success of this approach should be questioned.\\

We  present a  toy model  that  illustrates the
possibility  to reproduce the  observed ratios  for $p/\pi^+$  using a
model  to  incorporate  the  flow  in the existing  event generators. The aim is to demonstrate that
the  radial  flow,  as  claimed  in  hydro  calculations,  does  have  a
considerable  influence  on  $p/\pi^+$  ratios  in  a  wide  range  of
centralities starting from $pp$ collisions.  One might ask why 
are we   including the proton collisions in  these considerations -
knowing that  it is difficult to  expect flow in so  light a colliding
system.  We include  the  pp collisions  because  in the  conventional
radial flow analysis of  STAR ~\cite{adams}, making a simultaneous fit
to the slopes  of pion, kaon and proton  spectra an ``{\it equivalent
flow}''  of $\approx$  0.2c is found.  The  measured value  is  of course  much
smaller than the  one obtained in the most  central $Au+Au$ collisions
where the value of  flow reaches $\approx 0.6c $.\\

\noindent
The remainder of this work  is organized as follows: in section 2, we
give a brief description of the event generators, the  section 3 describes
our toy model of flow used to describe experimental
data.   In section  4, the results  of  our  model  and their  comparison  to
experimental  $Au+Au$ and  $p+p$ results  are presented.  Finally some
conclusion are drawn in section 5.

\section{The event generators}
The  main  tool  of  comparison  of the  measured  with  the  existing
knowledge is compiled  in the so called event  generators.  Among them
the  most prominent is  the PYTHIA\cite{Pythia63}  generator  for the
proton-proton collisions  and the HIJING~\cite{Hijing}  generator for the  heavy ion
collisions. We should  mention that  even proton-proton
collision can be  studied using HIJING. The version  6.2 of PYTHIA, does
not reproduce the proton  to pion ratio~\cite{star01} with its default
parameters. The possibility to  improve the situation by including the
Leading Order(LO)  and Next  to Leading Order  (NLO) corrections  which are
implemented in  PYTHIA by the  so called K-factor, has  been explored.
The  requirement of a  K-factor may  indicate collective  phenomena in
$pp$ collisions as in  heavy ions data~\cite{0606020}. The K-factor as
function of the energy has been extracted~\cite{Eskola05}  together
with  other phenomena as energy  loss for hard partons and temperature
effects to
explain the $p_t$ spectra. Those  studies indicate that while the pion
spectra can be  described with the default PYTHIA  settings, (i.e. QCD
processes at  leading order) the proton spectra  require the inclusion
of  a  K-factor~\cite{0606020}!!   (QCD  processes with  higher  order
corrections). Hence it is not possible to reach a consistent
reproduction of the experimental data.
The HIJING generator  dedicated to heavy ion reactions
does not reproduce the proton spectra in a similar way as PYTHIA.  One
has  to add  that neither  of  them includes  the radial  flow in  the
simulation,  an important  issue  to   describe  results  of  heavy  ions
collisions.

We use the  PYTHIA 6.3 generator with the popcorn
baryon production  mechanism.  One can change parameters  in the event
generator, like  the fragmentation function,  and/or the hadronization
mechanism.   Recently  it  has  been shown~\cite{cuautle05}  that  the
differences in the proton/pion ratio at 200 GeV among different baryon
production mechanisms are not very large so that we limit ourselves to
the use of the popcorn hadronization mechanism in the PYTHIA generator
- the  one that is  expected to  give the  best conditions  for proton
production.
The HIJING 1.32 event generator has been used to generate $Au+Au$ and pp
events, with default values of the parameters, for instance, including
partonic energy loss, shadowing effects, among other
phenomena. The pp collision with HIJING was generated with the default set
of parameters.\\

\section{Radial  flow: our model}

The radial flow is  understood as representing the azimuthally symmetric collective  aspects of the
interacting hadronic medium~\cite{NA44},  depending on the co\-lli\-sion energy.
The relevant observable to  study the radial
flow is the transverse momentum  of the particles.  For each particle,
the random  thermal motion is superimposed onto  the collective radial
flow velocity.  Consequently, the invariant $p_t$ distributions depends
on the temperature at freeze  out, the particle mass, and the velocity
profile of  the flow.   The experimental data  on radial flow  at RHIC
indicate that the kinetic  freeze  out  temperature and  the  observed flow  are
anti-correlated.  The temperature  decreases with centrality while the
flow velocity  increases~\cite{adams,cuautle05,Barannikova:2005rw}.
From the most peripheral to the most central the flow for heavy ions
collisions rises from $\approx$ $0.3c$ to $\approx$ $0.6c$.

\noindent
We propose to introduce radial flow, to event generators as follow: in
a  first step  we generate  {\it  flow-free} pion  and proton  spectra
using an  event generator to produce
particles.\\
We  are assuming that  a fireball,  thermalized, and  expanding was
created  in  the  collision.   The expansion  produces  an  additional
momentum to the one created  in the collisions using event generators.
This  contribution we  call momentum  of the radial flow $p_{t,f}$  given by
$p_{t,f}  =\gamma  m\beta$,  where  $\gamma$ is  the  Lorentz  factor,
$\beta$ is  the profile velocity and  $m$ is the mass  of the particle
under consideration.   This radial component is generated  in a random
way in the transverse plane and is added vectorially to the transverse
momenta  produced by  the generators,  supposed not  to  contain flow.
Once the  vector sum  is achieved, the  transverse momentum  $p_t$, of
each  particle generated  includes now  the radial  flow. Then  we can
select  the pions  and protons  and estimate  the ratio  comparing the
results with and without flow versus experimental data.\\
The radial flow described above  can be added to   any
event generator in order to compare among them and with the
experimental data.

\section{Results from our model versus data}

\begin{figure}[htb] 
{\centering
\resizebox*{0.85\textwidth}{0.45\textheight}
{\includegraphics{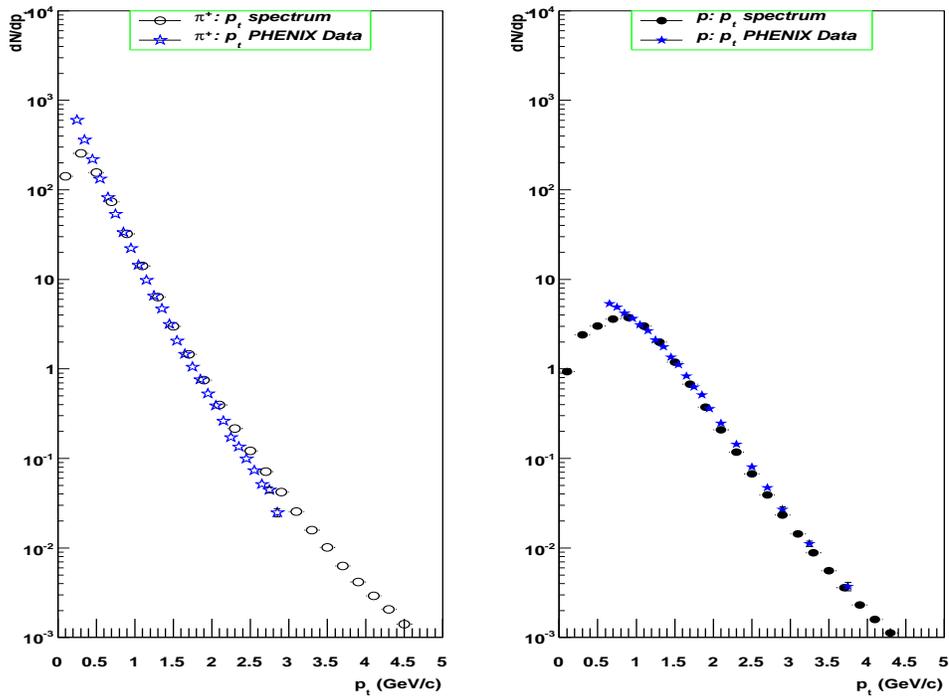}}
\par}
\caption{$p_t$ spectra  for pions (left) and  protons (right) obtained
applying the afterburner  to $Au+Au$ central collisions generated by HIJING with a flow
of $0.5 c$.}
\label{auau-pt}
\end{figure}

\begin{figure}[htb] 
{\centering
\resizebox*{0.45\textwidth}{0.35\textheight}
{\includegraphics{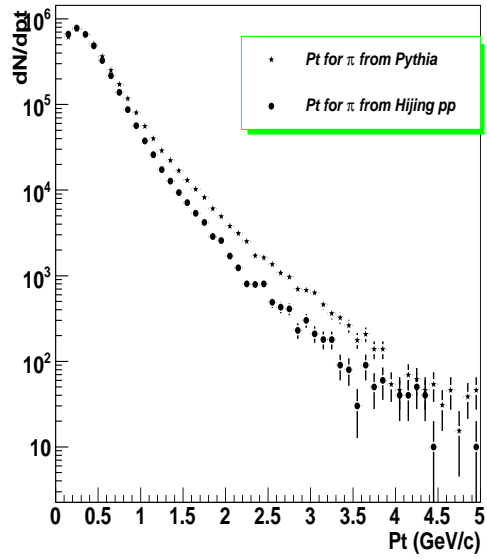}}
\par}
\caption{   $p_t$ spectra  of pions
as, generated for $pp$ collisions with PYTHIA and HIJING event generators.}
\label{pt-pp-auau}
\end{figure}

The flow  contribution considerably alters, as expected,  the shape of
the  momentum  spectra  in the  range  0  to  $\sim  4 $  GeV/c.   The
Fig.~\ref{auau-pt}  represents a typical  momentum spectrum  for pions
(left) and  protons (right), obtained for  central $Au+Au$ collisions
with  a velocity  profile of  the flow,  $\beta =  0.5c$. In  the same
figure  we show  the PHENIX  spectra~\cite{phenix} to show  the
degree of  agreement between the  spectra obtained with our  model and
data.  Spectra for $pp$ collisions using HIJING or PYTHIA shown in the
Fig.~\ref{pt-pp-auau}  indicate  that  there are  notable  differences
above  1~GeV/c,  between  the  two generators.   The  differences  are
probably  due to  the parton  distribution functions  used  and/or the
fragmentation functions.

\subsection{$Au+Au$ collisions at 200 GeV}
\begin{figure}[] 
\centering{
\resizebox*{0.45\textwidth}{0.35\textheight}
{\includegraphics{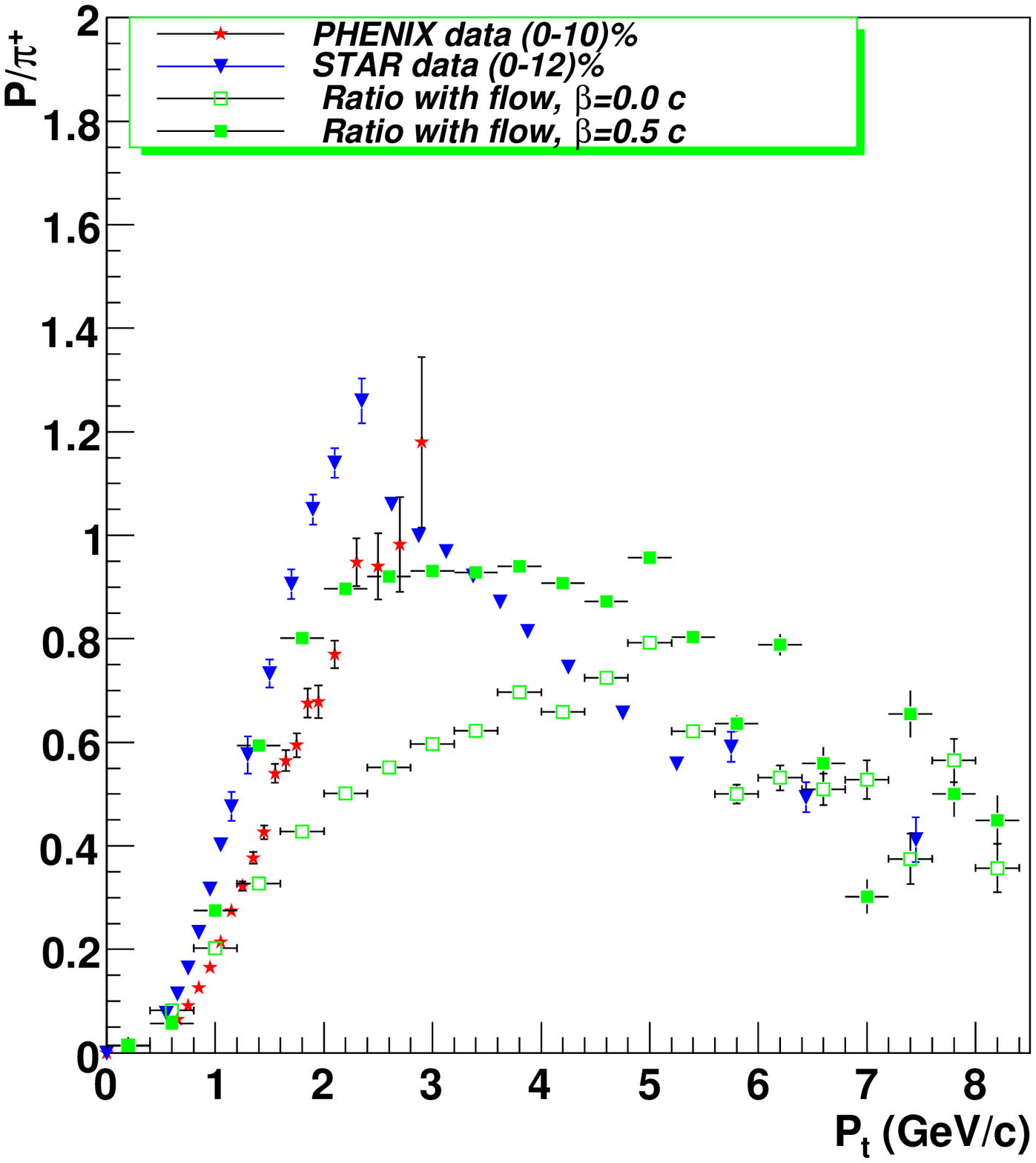}}
\par}
\caption{Proton to  pion ratios obtained  with the present  model with
and without  flow, compared to data  from PHENIX and STAR  for most
central collisions.}
\label{auau-0-10x100a}
\end{figure}

\begin{figure}[] 
\resizebox*{0.45\textwidth}{0.35\textheight}
{\includegraphics{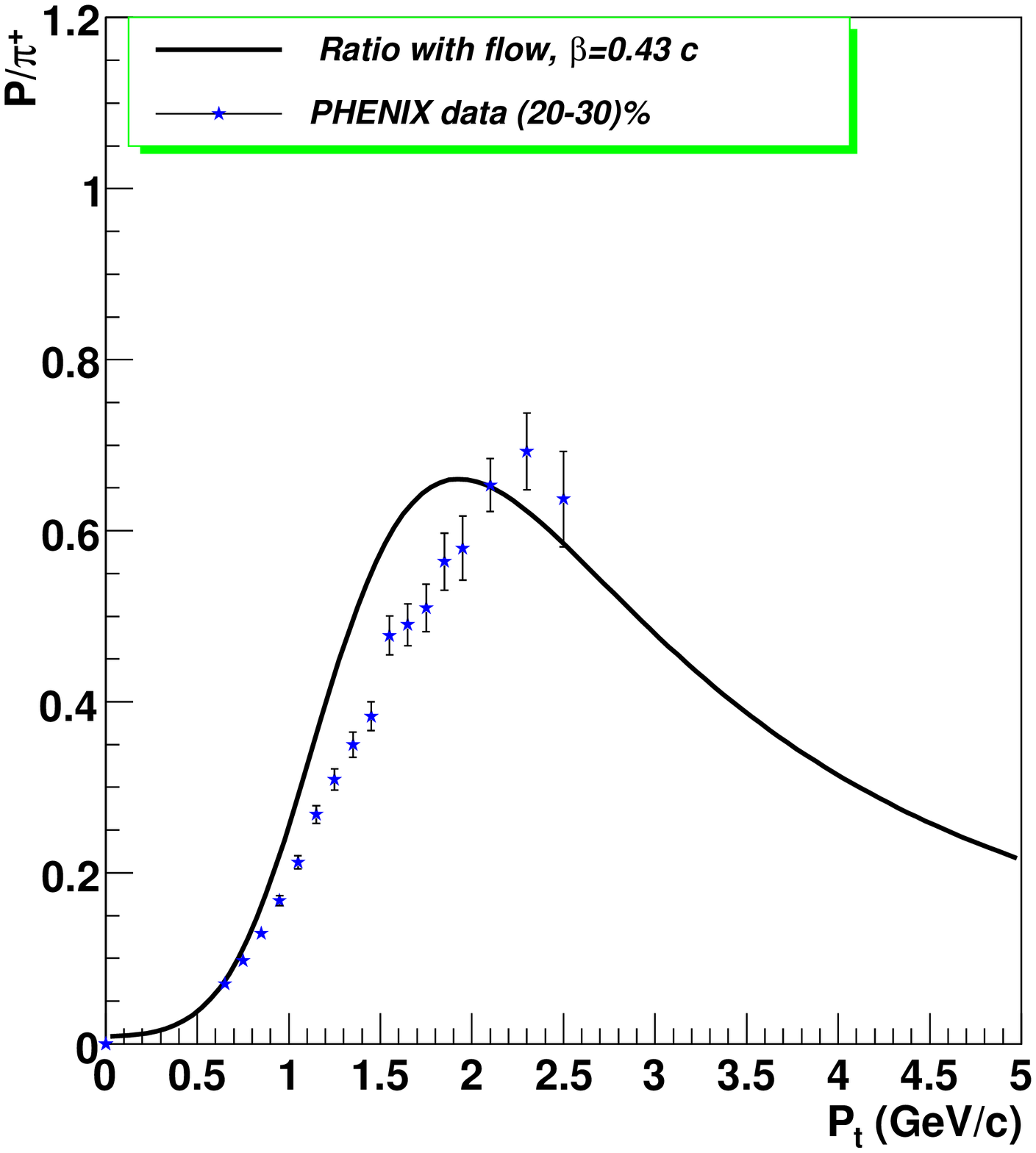}}
\resizebox*{0.45\textwidth}{0.35\textheight}
{\includegraphics{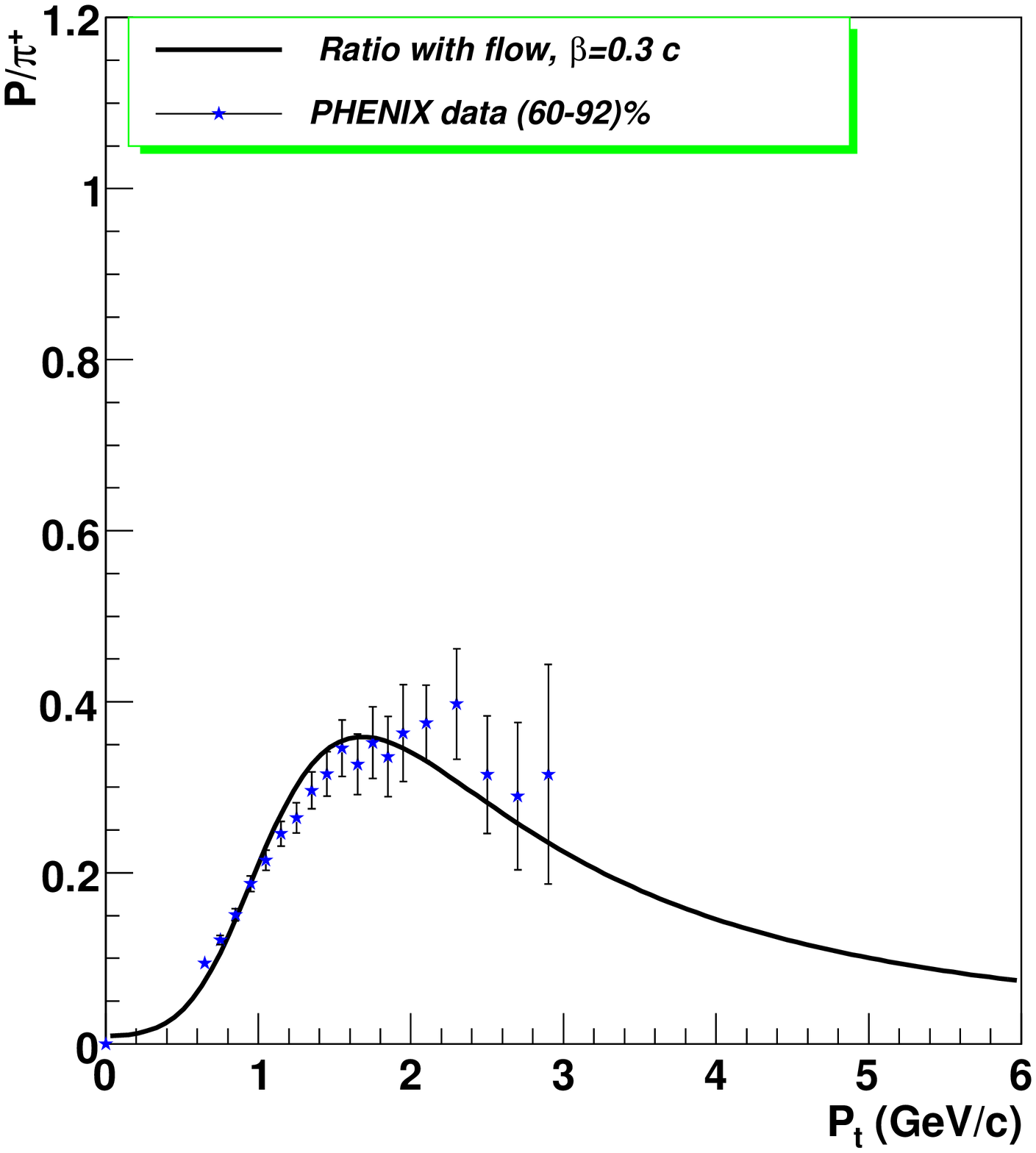}}
\caption{Proton  to  pion  ratios  obtained  with  the  present  model
compared to  data at  two different centralities.   The data  are from
PHENIX. The  solid lines were obtained  as the ratio of  the fits to
the simulated proton and pion spectra.}
\label{auau-0-10x100b}
\end{figure}

\noindent
Fig.~\ref{auau-0-10x100a}  shows   the  results  of   our  calculation
 including the flow in the generator, for the proton to pion ratio for
 $Au+Au$   collisions,   compared   to   two   experimental   results,
 PHENIX~\cite{phenix}  and   STAR~\cite{Abelev:2007ra}  for  the  most
 central collision.  We also plotted  the results  without
 flow ($\beta =  0.0c$), to illustrate the importance  of the flow the
 contribution to the ratio. The other centralities are also reasonably
 described  using lower  flow parameters,  following  the experimental
 results,  as shown in  Fig.~\ref{auau-0-10x100b}.  Without  trying to
 get the best fit, the distributions show a qualitative agreement with
 the experimental results with a rise and a subsequent decrease of the
 ratio at $p_t$  values from $\approx 2.5 - 3$  GeV/c onwards, for the
 three different centralities.

\noindent


\subsection{$Au+Au$ collisions at 62.4 GeV }

\begin{figure}[] 
{\centering
\resizebox*{0.45\textwidth}{0.35\textheight}
{\includegraphics{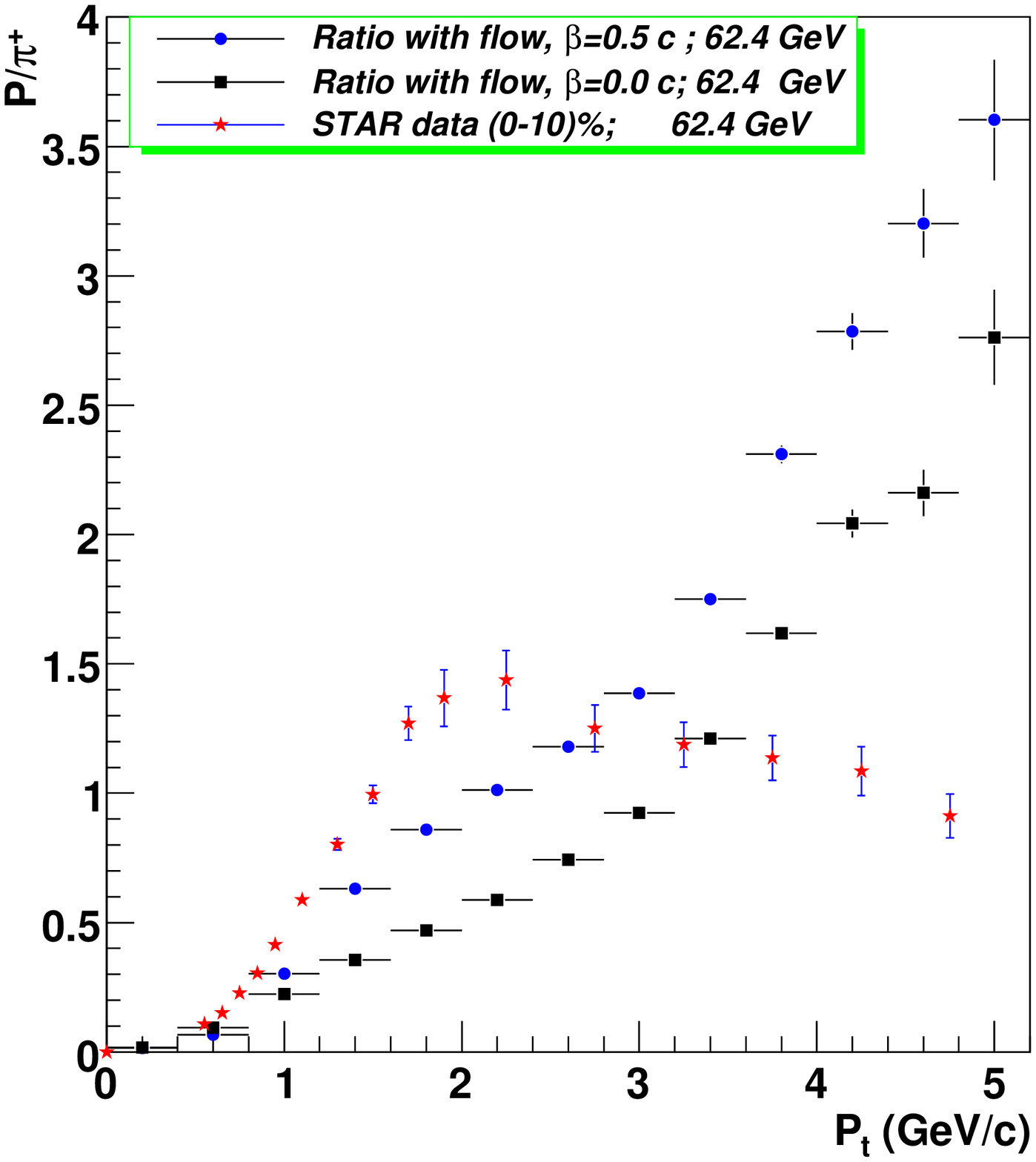}}
\resizebox*{0.45\textwidth}{0.35\textheight}
{\includegraphics{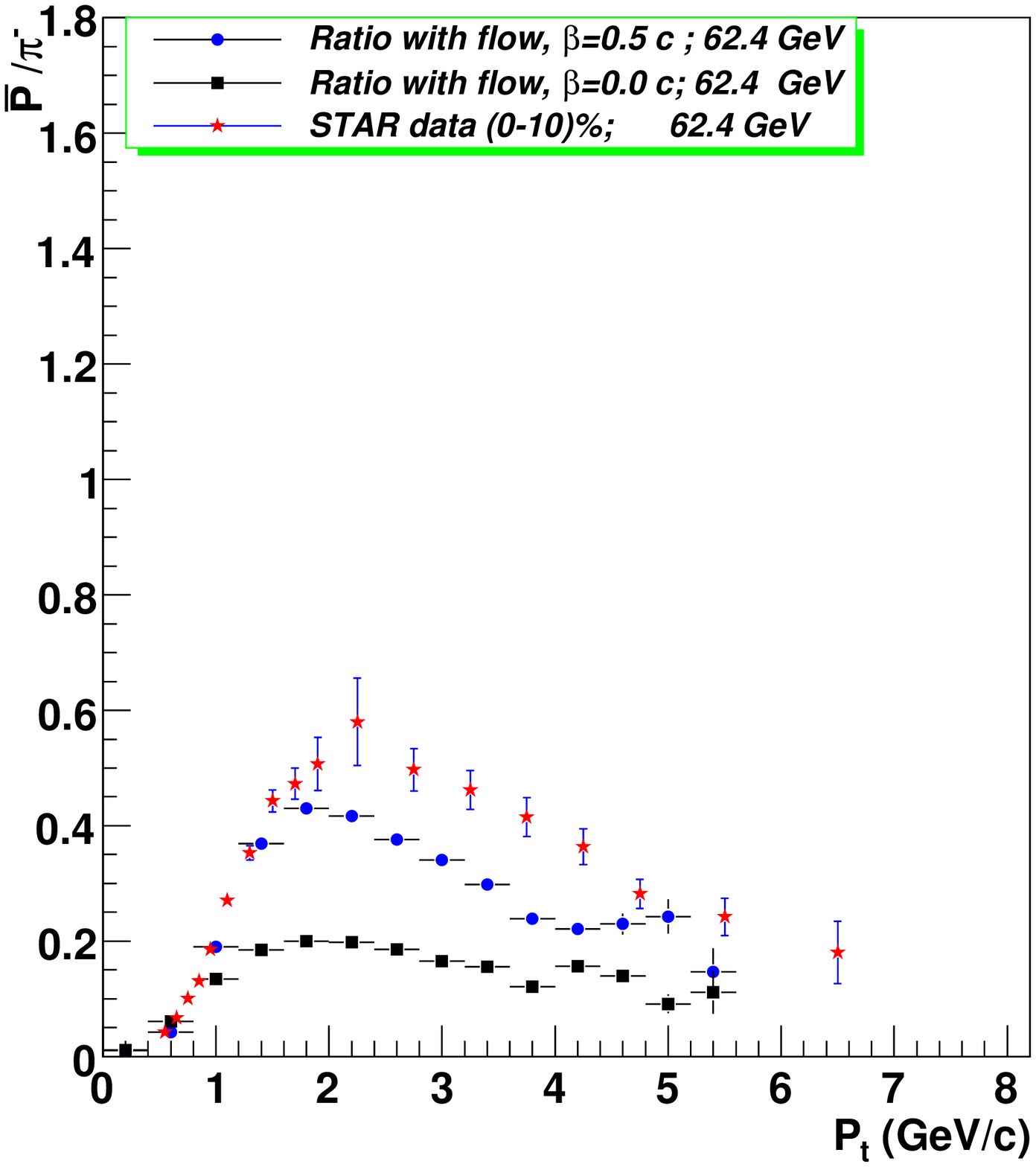}}
\par}
\caption{Proton to pion ratio  from our model compared to central $Au+Au$
data at 62.4 GeV. The left part shows the results for $p/\pi^+$ while
the right part shows the $\bar p/ \pi^-$ ratio.}
\label{auau-0-10-62}
\end{figure}

The  data for  protons  at 62.4  GeV  are difficult  to analyze  using
HIJING.  The  HIJING generator  at this energy,  namely, overestimates
grossly the proton  production with respect to pions.  This is due, in
our opinion, to the fact that HIJING overestimates the contribution of
valence quarks at these momenta thus grossly overestimating the proton
production  compared to  the antiproton  one, in  the same  way  as it
underestimates the  $\bar p/p$  ratio at these  momenta.  In  the left
part of the Fig.~\ref{auau-0-10-62}, we show the results of the HIJING
predictions  of  $p/\pi$ compared  with  and  without the  afterburner
versus data.  The results are completely different  for anti-proton to
pion ratio.  The right part  of the Fig.~\ref{auau-0-10-62},  show the
$\bar  p  /\pi^-$ data  of  STAR\cite{Abelev:2007ra}  with the  HIJING
predictions with and without afterburner.   It is visible that the two
predictions are  completely different  and that the  afterburner again
qualitatively  reproduces the  data although  they suggests  perhaps a
somewhat larger flow.
Let us note that the trend of our model is to predict the position of
the maximum ratio at lower momenta than at 200 GeV in agreement with
the data, while the coalescence models used
in ref~\cite{Abelev:2007ra} show an opposite behavior.

\subsection{$pp$ collisions}
\begin{figure}[] 
{\centering
\resizebox*{0.45\textwidth}{0.35\textheight}
{\includegraphics{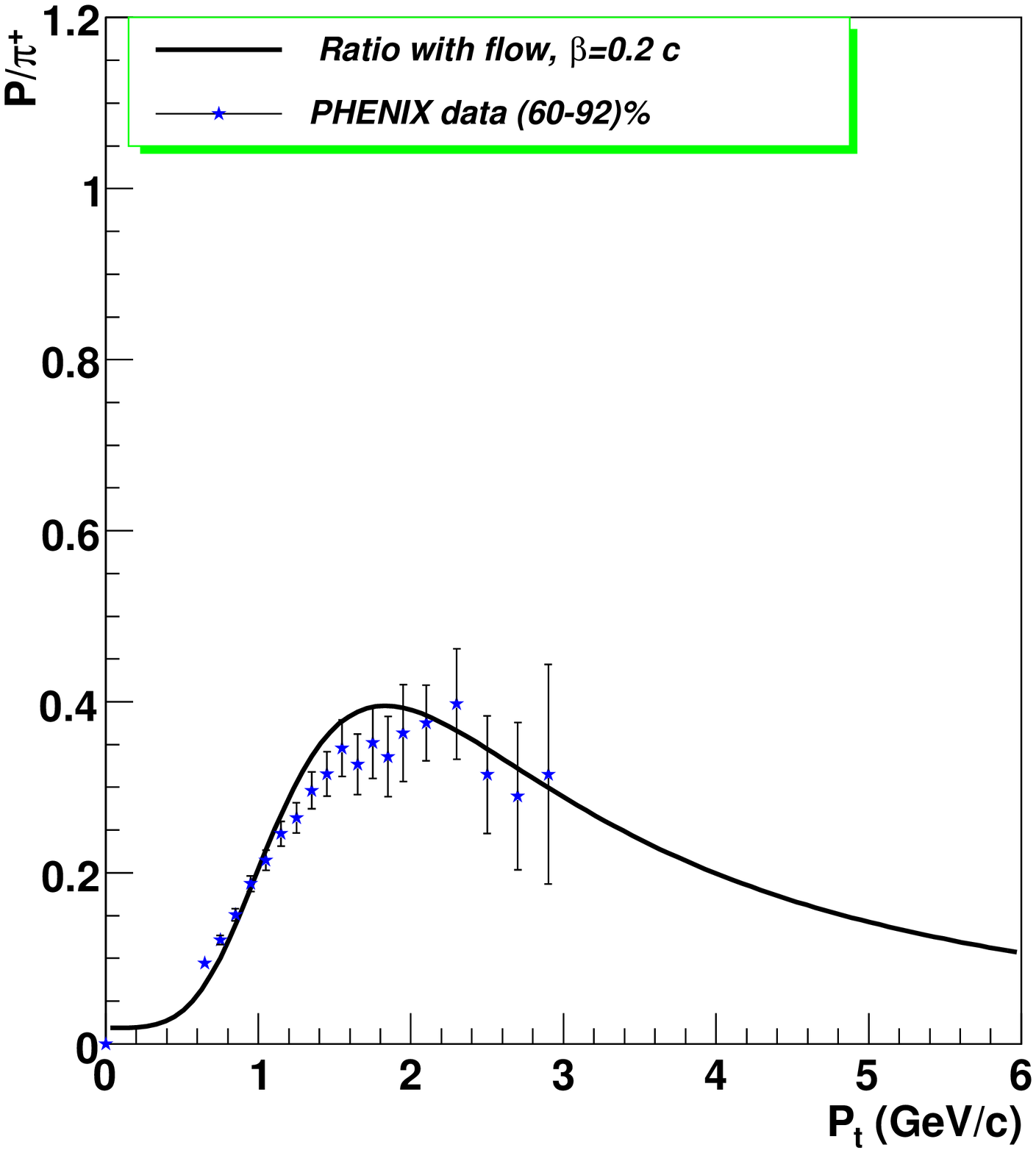}}
\resizebox*{0.45\textwidth}{0.35\textheight}
{\includegraphics{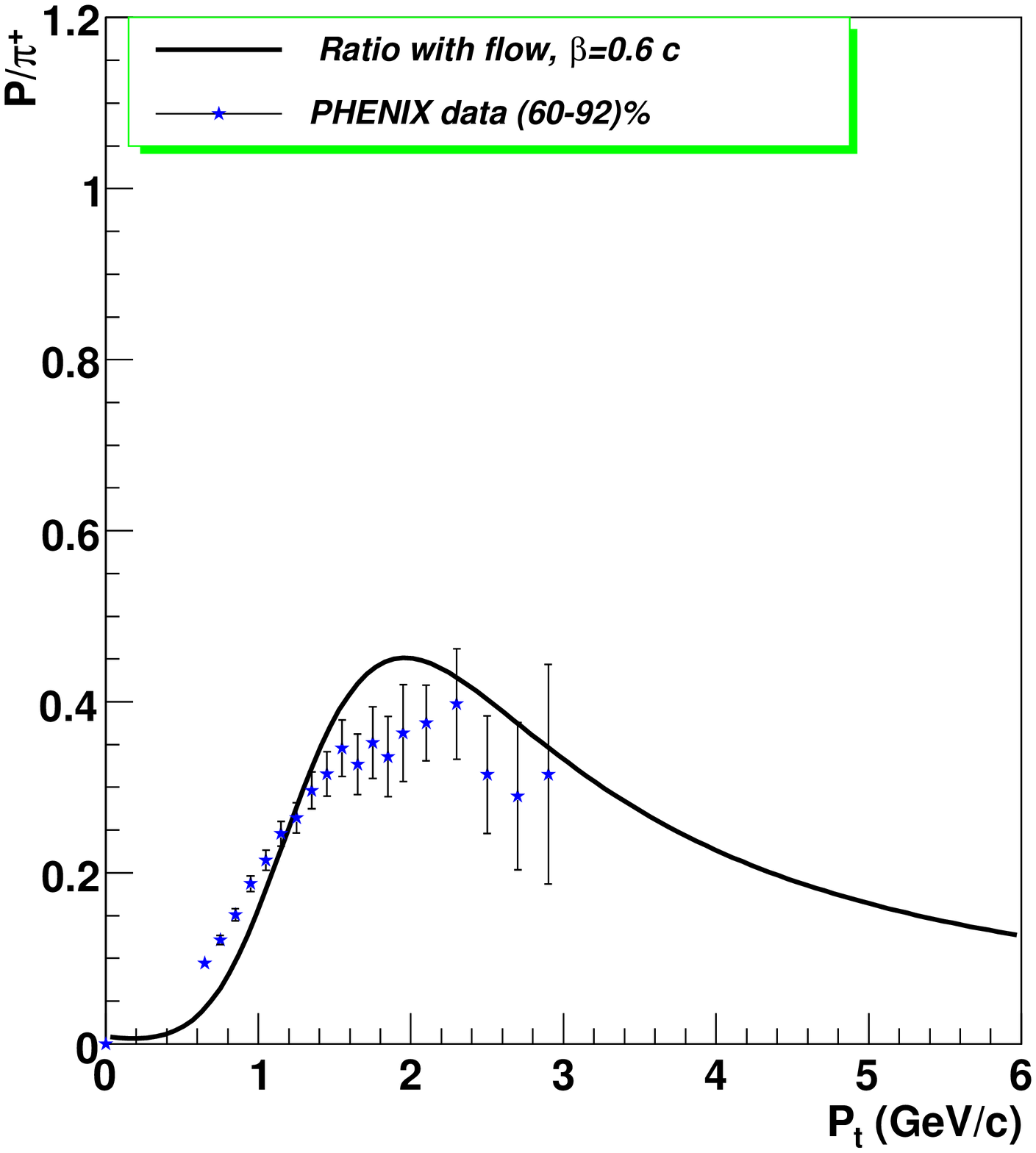}}
\par }
\caption{Proton  to  pion ratio  from  HIJING  compared to  peripheral
$Au+Au$ data (left). The  same data compared with PYTHIA generation
and flow $0.6 c$ is shown in the  right. The solid lines were obtained
as the ratio of the fits to the simulated proton and pion spectra.}
\label{auau-60-92x100-1}
\end{figure}

The fact  that even the ratios  obtained in $pp$  collisions cannot be
satisfactorily  explained  with  current  generators as  discussed  by~\cite{star01},  prompted us  to apply our  model to
them since the  experimental analysis of the spectra  in $pp$ yields a
value of 0.2c  like in the case of  peripheral $Au+Au$ collisions.  We
show in Fig.~\ref{auau-60-92x100-1} the  results obtained for the very
peripheral  - quasi $pp$  like collisions  using our afterburner one
with HIJING (left) and one with PYTHIA  (right). We find that PYTHIA requires a
very large  flow to  fit the  data while HIJING  with a  moderate flow
similar  to the one  extracted from  the experiments,  fits reasonably
well.   This  illustrates that  the  ratio  depends  crucially on  the
"initial   spectrum",  as  discussed   in  section   2  (fragmentation
functions, parton distribution functions, etc.).  Also it should be mentioned
that  the HIJING event  generator implements the  nuclear effects
like shadowing, energy loss, etc.  even in peripheral collisions.

\begin{figure}[] 
{\centering
\resizebox*{0.45\textwidth}{0.35\textheight}
{\includegraphics{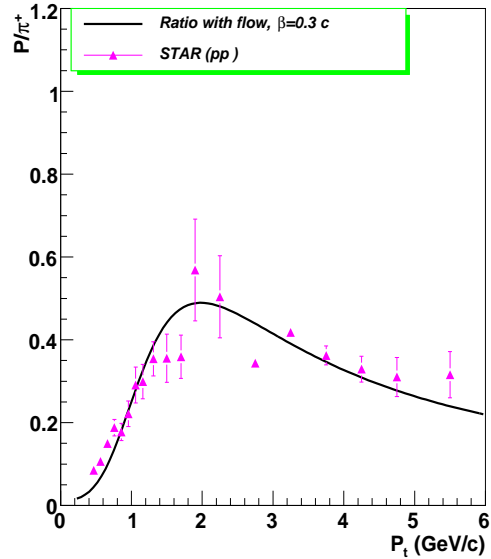}}
\par}
\caption{Proton  to  pion ratio  from  $pp$  collisions using  HIJING,
  compared  to $pp$  data from  STAR  experiment. The  solid line  was
  obtained as  the ratio of  the fits to  to the simulated  proton and
  pion spectra.}
\label{auau-60-92x100-2}
\end{figure}

\noindent
Results  from  STAR~\cite{star01}   $pp$  collisions  have  been
compared with  HIJING events.  In  Fig.~\ref{auau-60-92x100-2} we show
this ratio.   As in the  case of peripheral  data a good  agreement is
achieved incorporating  the radial flow to the  HIJING spectra, albeit
the flow used is somewhat larger than the one found experimentally.

\noindent

\section {Conclusions}
We have  introduced radial flow  in existing event  generators (PYTHIA
 and  HIJING)  via  an  afterburner.   The flow  parameters  that  fit
 reasonably well the  experimental results both in the proton to pion
 ratio and the $p_t$ spectra are very close  to the ones
 extracted  from  experiments  in  a thermal  analysis.   The success in
 reproducing qualitatively  the data in a wide  range of centralities,
 and  in a  large  range  of transverse  momentum  indicates that  the
 presence  of radial  flow should  be  taken in  consideration in  any
 attempt to explain the so called proton puzzle.

In the present work we did  not try to reproduce in details the ratios
experimentally  observed for  two reasons;  first the systematical and
statistical errors of the experimental results and   second  as   has  been   shown  the
``initial condition'' of the generator  have influences on the shape of
the spectra leads to  difference in the results after implementation
of  the
radial flow.  \\ The inclusion  of a flow  in pp reactions  is not
justified  in terms  of an  expanding  thermal system,  but since  the
analysis of the  pion kaon and proton spectra  do yield something that
may be termed ''flow'' we can perhaps infer that the outward motion of
jets produces  an effect  which does modify  the hadron spectra  in pp
collisions.

\section{Acknowledgments} 
We would  like to thank A.  Morsch for helpful  discussions. This work
was supported  in part  by PAPIIT project  IN116508 and  CONACyT under
grant numbers 52162-F, 40025-F, 44268-F .

\end{document}